\documentstyle[12pt]{article}

\newcommand \cN  {{\cal N}}
\newcommand \ext {\mathop{\rm Ext}}
\newcommand \oh {{\textstyle\frac{1}{2}}}
\newcommand \oq {{\textstyle\frac{1}{4}}}
\newcommand \oeight {{\textstyle\frac{1}{8}}}
\begin{document}

\title{On the number of metastable states\\
       in spin glasses}
\author{Giorgio Parisi
        and Marc Potters\thanks{\tt potters@roma1.infn.it}\\[0.5em]
  {\small Dipartimento di Fisica and INFN, Universit\`a di Roma
    {\em La Sapienza}}\\
  {\small \ \  P. A. Moro 2, 00185 Roma (Italy)}\\[0.5em]}
\date{\today}
\maketitle

\begin{abstract}
  In this letter, we show that the formulae of Bray and Moore for the
  average logarithm of the number of metastable states in spin
  glasses can be obtained by calculating the partition function with
  $m$ coupled replicas with the symmetry among these explicitly broken
  according to a generalization of the `two-group' ansatz. This
  equivalence allows us to find solutions of the {\sc bm} equations
  where the lower `band-edge' free energy equals the standard static
  free energy. We present these results for the
  Sherrington-Kirkpatrick model, but we expect them to apply to all
  mean-field spin glasses.
\end{abstract}

\vfill

\begin{flushright}
  {\bf  cond-mat/9506049}
\end{flushright}

\newpage

\noindent
There has been recently a renewed interest in the study of metastable states
in spin glasses.  Current problems range form their potential relevance in
spin glass dynamics to the formulation of renormalization group analysis in
the presence of many minimas of the action. Nevertheless, it is
interesting to use our present knowledge to shed some light old problems
and to understand the connection between new and older techniques.

One recent development is the use of coupled real replicas as a way to
explicitly break ergodicity revealing the `complexity' of the system
\cite{MONASSON,FRAPAR}.  It was not known up to now whether or not
these techniques were equivalent to the now classical computations of
the number of {\sc tap} solutions of Bray and Moore ({\sc bm})
\cite{BM,BMRSB}.  It has been shown to be true for the spherical
$p$-spin model, the Potts glass and for part of the spectrum of the
random orthogonal model ({\sc rom}) \cite{CRISOM,KIRWOL,PARPOT}.  A
much older question that remained partly unanswered is whether the
computation of {\sc bm} for the Sherrington-Kirkpatrick model ({\sc
  sk}) using continuous {\sc rsb} gives at the lower `band-edge' the
same free energy as the static calculation.  It was shown in Ref.\
\cite{BMRSB} that they agreed near the transition, at least to the
first non-trivial order in the reduced temperature $t$.  The {\sc bm}
computation involves a function $Q(x)$ similar to the usual Parisi
function $q(x)$ but in this computation the critical value of $x$
($x_{\rm c}$) is equal to $\oh$ near $T_{\rm c}$ while its static
counterpart only equals $2t$.  Finally, a never explained coincidence
is the fact that the `direct average' of {\sc bm} both for {\sc sk}
and {\sc rom} can be obtained by calculating the partition function
under the two-group replica symmetry-breaking scheme.  If one does not
let the number of replicas $n$ go to zero, the two-group partition
function gives the Laplace transform of the number of {\sc tap}
solution with $n$ conjugated to $-\beta f$ \cite{PARPOT,BM2G2}.

The contribution of this letter is to show that a generalization of the
two-group ansatz makes the connection between the `real replicas' and the
{\sc bm} framework. This connection in turns helps us understand why, in
the `naive' two-group approach, $n$ becomes the Laplace conjugate of the
free energy. The two-group model, in its unbroken form, gives us the
proper ansatz to look for solutions of the {\sc bm} saddle point that
reduces to the static free energy at the lower band-edge. All these
points are illustrated on the {\sc sk} model but we believe that they
should apply to any mean-field spin glass. Unfortunately, a crucial point
stills eludes us, that is the physical significance of the two-group
ansatz and the difference, if any, between metastable states that break
the two-group symmetry and those that don't.

In Ref.\ \cite{MONASSON} it was argued that the number of pure states
in a glassy system can be computed by introducing $m$ real replicas of
the system. More precisely, for a mean-field model with quenched
randomness where one introduces $n$ replicas to average over the
disorder, we have the following relation
\begin{equation}
\label{E_BASIC}
\max_{f} \left[\left\langle \log \cN(f)\right\rangle -m\beta f\right]=
\lim_{n\rightarrow 0} \frac{1}{n} \log \left\langle Z^{mn}\right\rangle,
\end{equation}
where $\cN(f)$ is the number of metastable states as a function of
their free energy and $Z^{mn}$ is the partition function of $n\times
m$ copies of the original system with the replica symmetry explicitly
broken into $n$ groups of $m$ replicas.

It is useful to consider both $f$ and $\langle\log\cN(f)\rangle$ as
function of the independent parameter $m$ through the relation
$f=f^{*}(m)$ i.e.\ the particular $f$ that maximizes the l.h.s.\ of
Eq.\ (\ref{E_BASIC}).  Varying $m$ then traces out the
$\langle\log\cN(f)\rangle$ vs $f$ curve, in general a bell shaped
curve (e.g.\ {\sc sk} or {\sc rom} \cite{BM,PARPOT}) or an
`interrupted bell' shape (e.g.\ $p$-spin, both spherical and Ising
\cite{CRISOM,RIEGER}). The maximum of this curve is given by $m=0$
(white average) while increasing $m$ favors states of lower free
energy. In particular $m=1$ picks out the states that contribute to
the Gibbs measure. Finally, we call the lower band-edge the infimum of
the free energies for which $\langle\log\cN(f)\rangle > 0$.

This formula along with the simplest replica symmetry ansatz gives a
very good description of the metastable states for mean-field models
with `first-order {\sc rsb}' (e.g.\ spherical $p$-spin, the Potts
glass and {\sc rom}).  It predicts that at the dynamical transition, a
large number of metastable states start to contribute to the free
energy (extensive complexity) and that at the {\sc rsb} transition the
complexity ceases to be extensive \cite{MONASSON}.  In the case of
{\sc sk}, a simple one-step ansatz in Eq.\ (\ref{E_BASIC}) always
gives either zero or a negative result for the complexity. As we shall
see, the presence of metastable states in {\sc sk} can be uncovered by
a more complex replica symmetry breaking scheme.

The partition function of the replicated {\sc sk} model is given by
(see e.g.\ \cite{BEYOND})
\begin{equation}\label{E_PART}
\left\langle Z^{mn}\right\rangle= \ext_{\bf Q}\exp \left\{N\left[
\frac{\beta^2}{4}\left(mn-{\rm Tr}{\bf Q}^2\right)
+\log \sum_{\{\sigma_a^i\}}
\exp\left( \oh\beta^2\sum_{abij}Q_{ab}^{ij}\sigma_a^i\sigma_b^j\right)
\right]\right\},
\end{equation}
where the extremum is taken over matrices $Q_{ab}^{ij}$ ($1\leq
a,b\leq n$ and $1\leq i,j\leq m$) with zeros on the diagonal
$Q_{aa}^{ii}\equiv 0$.

The point of this letter is to show that the formula for the average
logarithm of the number of {\sc tap} solution of Ref.\ \cite{BMRSB}
can be obtained from Eq.\ (\ref{E_BASIC}) using a direct generalization
of the `two-group' replica symmetry breaking \cite{BM2G2}.
We will use the following ansatz on each of the $n^2$ sub-matrices
${\bf Q}_{ab}$ of size $m\times m$ (i.e.\ the matrices $Q_{ab}^{ij}$ with
$(a,b)$ fixed and $(i,j)$ varying form 1 to $m$):
\begin{equation}
{{\bf Q}_{ab}}=
\left(\mbox{\rule[-1.5ex]{0cm}{3ex}}\right.
    \overbrace{
      \begin{array}{c}
         Q^+_{ab}\\
         Q_{ab}
       \end{array}
      }^{y}
     \:
    \overbrace{
      \begin{array}{c}
         Q_{ab}\\
         Q^-_{ab}
       \end{array}
      }^{m-y}
      \left.\mbox{\rule[-1.5ex]{0cm}{3ex}}\right) ,
\end{equation}
where $y$ is to be taken to infinity. A convenient way to parameterize
the matrices $Q^+_{ab}$ and $Q^-_{ab}$ in order to obtain a
finite result as $y\rightarrow\infty$ is
\begin{equation}\label{E_PARA}
Q^{\pm}_{ab}=Q_{ab}\pm\frac{A_{ab}}{y}+\frac{B_{ab}}{y^2}\,.
\end{equation}
Note that the `diagonal matrices' ${\bf Q}_{aa}$ have the extra
constraint that their diagonal elements $Q_{aa}^{ii}$ be zero; their
off-diagonal elements, in general non-zero, are also parameterized
according to Eq.\ (\ref{E_PARA}). We do not need to specify the
structure of the three matrices $Q_{ab}$, $A_{ab}$ and $B_{ab}$ in
order to make the connection with the {\sc bm} formulation, but in
practice they will be taken to have the usual utrametric structure
(see e.g.\ \cite{BEYOND}).  Using this ansatz, we can compute the
partition function (\ref{E_PART}) in a manner similar to that of Ref.\
\cite{PARPOT}. We find
\begin{eqnarray}\nonumber
\left\langle Z^{mn}\right\rangle
&=&\ext\exp\left\{nN\left[m \log 2 +\oq\beta^2 m(1-Q)^2
-\beta^2 A(1-Q)\right]\right.\\
&&\left.-N\beta^2\left[\sum_{ab} \oh A_{ab}^2
+Q_{ab}B_{ab}+mQ_{ab}A_{ab}+\oq m^2Q_{ab}^2\right]+N \log I\right\},
\label{E_ZMN}
\end{eqnarray}
where
\begin{eqnarray}\nonumber
I&=&
{\int_{-\mbox{\scriptsize i}\infty}^{\mbox{\scriptsize i}\infty}}
\prod_{a}\frac{dx_a}{2\pi{\rm i}}\int_{-1}^{1}\prod_{a}\frac{dm_a}{(1-m_a^2)}
\exp\left\{-\sum_{a} \left[x_a\tanh^{-1}m_a \right.\right.\\
&&\left.+\oh m\log(1-m_a^2)\right]
\left.+\beta^2 \sum_{ab}\oh Q_{ab}x_a x_b+B_{ab}m_a m_b+A_{ab}m_a x_b\right\},
\end{eqnarray}
and where $A$ and $Q$ are the diagonal terms $A_{aa}$ and $Q_{aa}$,
note that the sum over $(a,b)$ also includes diagonal terms.
If we put this formula in Eq.\ (\ref{E_BASIC}), we find that the
average log number of metastable states is identical to the average
log number of {\sc tap} solutions as computed in Ref.\ \cite{BMRSB}.
We recall that in that paper, the number of {\sc tap} solutions is
written as the integral over all configurations of an appropriate
delta function, replicas are introduced to average the logarithm over
the disorder. The equivalence between our formula and Eq.\ (11) of
Ref.\ \cite{BMRSB} can be rendered more obvious by the
following change of variables,
\begin{displaymath}
\eta_{ab}=\beta^2 Q_{ab}
\,,\;
\rho_{ab}=-\beta^2 \left(A_{ab}+\oh mQ_{ab}\right)
\,,\;
\eta^{*}_{ab}=\beta^2\left(2B_{ab}+mA_{ab}+\oq m^2 Q_{ab}\right)
\,,
\end{displaymath}
\begin{equation}\label{E_CHANGE}
q=Q
\,,\;
\Delta=\beta^2 \left(A+\oh mQ\right)
\,, \;
\lambda=\beta^2\left(B+\oh mA+\oeight m^2 Q\right)
\;\mbox{and}\;
u=-m\,,
\end{equation}
and a shift of the integration variable ($x'_a=-x_a+\oh m m_a$).

In Ref.\ \cite{BM2G2} it was shown that the original two-group
partition function (with $n$ in general non-zero) gives the Laplace
transform of the annealed averaged number of metastable states with
$n$ conjugated to $-\beta f$. We recover this annealed result by
setting the `off-diagonal' matrices ${\bf Q}_{ab}$ to zero, leaving us
only with $n$ diagonal blocks of $m\times m$ two-group matrices. This
explains the strange scaling with $n$, i.e.\ the size of the total
matrix $Q_{ab}^{ij}$ has to be taken to zero, while the $ij$ block
size $m$ remains finite and becomes the Laplace conjugate of the free
energy.  Similarly, using an $ij$-two-group over an {\sc rs} $Q_{ab}$,
we recover the `innocent replica' treatment, in this case, the full
matrix $Q^{ij}_{ab}$ can be viewed as a one-step matrix with blocks of
size $m$ within which lies a two-group structure.

Our formulation of the {\sc bm} problem allows us to find a class of
simple solutions to the saddle point equations, i.e.\ the `unbroken
two-group'. Indeed one can show that $A_{ab}=B_{ab}=0$ forms
self-consistent solution of these equations.  The problem is then
reduced to finding the extremum with respect to a single matrix
$Q_{ab}$ with a diagonal term $Q$. The computation with a $k$-step
{\sc rsb} ansatz on the matrix $Q_{ab}$ is equivalent to the usual
computation of the partition function with $k+1$-steps with {\sc rsb}
parameters $0\leq mx_1 \leq\ldots\leq mx_k \leq m \leq 1$. In
particular, since Eq.\ (\ref{E_BASIC}) implies
\begin{equation}
  \log \left\langle\cN(f)\right\rangle =-m^2 \frac{dF(m)}{dm} \mbox{
    when } \beta f=\frac{d}{dm} (mF(m))\,,
\end{equation}
where
\begin{equation}
  F(m)=\lim_{n\to 0} \frac{1}{mn}\log \left\langle
  Z^{mn}\right\rangle,
\end{equation}
the condition for the lower band-edge is just the usual minimization
with respect to the last {\sc rsb} parameter $m$.

In the simplest case, the `innocent replicas' admit a solution that is
equivalent to the one-step partition function. We have explicitly
checked that the expansion of (\ref{E_ZMN}) near $T_{\rm c}$ (Eqs.\
(12,13)\footnote{We think that the factor $(1-4Q)$ in that second
  equation should read $(1-5Q)$.} of Ref.\ \cite{BMRSB}) admits a
saddle point of the form (\ref{E_CHANGE}) with ${\bf A}={\bf B}=0$, we
find, for the lower band-edge
\begin{displaymath}
q=t+{\textstyle \frac{25}{27}}t^2+{\rm O}(t^3)
\,,\;
q_{\circ}={\textstyle \frac{1}{3}} t +{\rm O}(t^2)
\,,\;
m={\textstyle \frac{4}{3}} t +{\rm O}(t^2),
\end{displaymath}
and
\begin{equation}
\beta f/N=-\log 2 -\oq \beta^2+{\textstyle \frac{1}{6}} t^3+
{\textstyle \frac{11}{24}} t^4+
(5099/7290) t^5 +{\rm O}(t^6)\,.
\end{equation}
This solution should be compared with that of Ref.\ \cite{BM}, where
it was found $q_\circ=0.3471 t +{\rm O}(t^2)$.  Our solution can be
followed to other values of the parameter $m$, but we find it to be
only valid at the lower band-edge.  As expected, for $m>\frac{4}{3}
t$, it gives a negative log-number of {\sc tap} solutions. Whereas for
$m<\frac{4}{3} t$, $f$ decreases with decreasing $m$ indicating that
we are on an unphysical branch. The correct solution for
$m<\frac{4}{3} t$ must therefore include two-group symmetry breaking.

The same `unbroken two-group solution' can be found for continuous
{\sc rsb} near $T_{\rm c}$. We find a solution very similar to that of Ref.\
\cite{BMRSB}, except for the value of $m$ which in turns changes the
value of all other parameters. We find
\begin{displaymath}
  q=t+t^2+{\rm O}(t^3) \,,\; q(x)=t x \: (x\leq x_{\rm c}=1) \,,\; m=2t
  +{\rm O}(t^2)\,,
\end{displaymath}
and
\begin{equation}
  \beta f/N=-\log 2 -\oq \beta^2+{\textstyle \frac{1}{6}} t^3+
  {\textstyle \frac{11}{24}} t^4+ {\textstyle \frac{7}{10}}t^5 +{\rm
    O}(t^6)
\end{equation}
The apparent discrepancy between the value of $x_{\rm c}=1$ and that of the
usual static solution ($x_{\rm c}=2t$) is understood by realizing that it is
not the matrix $Q_{ab}$ which should equal the static one but rather
the full matrix $Q^{ij}_{ab}$. When $n$ goes to $0$, the full matrix
can be parameterized by a function $q_{\rm full}(x)$,
\begin{equation}
q_{\rm full}(x)=q(x/m)\,,\; 0 \leq x\leq m \, ; \;\;
q_{\rm full}(x)=Q     \,,\; m<x\leq 1\,,
\end{equation}
in perfect agreement with the static solution.

\subsection*{Acknowledgments}
We wish to thank M. Ferrero and R. Monasson for useful discussions.

\end{document}